\documentclass[preprint,prx,longbibliography]{revtex4-1}
\usepackage{subcaption}
\usepackage{graphicx}
\usepackage{comment}
\usepackage{amsmath,amssymb,amsthm} 
\usepackage{scalerel,stackengine}
\usepackage{bm}
\usepackage{verbatim}
\usepackage{float}
\usepackage{color,soul}
\usepackage{textcomp}
\usepackage{rotating}
\usepackage{color,soul}
\usepackage{mathrsfs}
\usepackage[title]{appendix}
\usepackage{xcolor}
\usepackage{mhchem}
\begin{document}
\title [mode = title] { Stability of Homochiral Peptides in Capillary Environments: Implications for the Origin of Biological Homochirality}
\author{S\o ren Toxvaerd}
\affiliation{ DNRF centre  ``Glass and Time'', Department of Science and Environment, Roskilde University, Post Box 260, DK-4000, Denmark}
\email{Submittet to BioSystems.  Email: st@ruc.dk}
\begin{abstract}
The homochiral proteins in living organisms are rather stable, but they tend  to lose
their secondary conformations and homochiral order over time.
This instability is found in all living organisms, from bacteria to humans. The  aging
of living organisms, with the destruction of their vital homochiral order, raises the  questions
about the emergence of homochirality and how the homochiral order in the building molecules was ensured in  prebiotic
times before life evolved. Based on thermodynamic arguments, we propose that the homochiral
order was achieved by a  ``Frank-synthesis'' of  homochiral peptides from racemic amino acids and in a capillary environment
in the vents in the Earth's crust, perhaps at crystalline chiral surfaces, and that the peptides
are stable over long times in such an environment.  The $hypothesis$ can be confirmed by experimental analysis of
the stability of the secondary structures in proteins as a function of water
activity and in a capillary environment.	
\end{abstract}
\maketitle
\section{Introduction}
All living organisms are with homochiral peptides consisting of units of L-amino acids, and carbohydrates with D-carbohydrate units.
These configurations are obtained by  asymmetric synthesis, and several questions  arise, such as:
When were these homochiral configurational structures synthesised,  how, and where?

When the asymmetric synthesis took place can be determined roughly, since the oldest forms
of life we know with reasonable certainty are  colonies of bacteria about
 3.5–3.8 billion years old \cite{Djokic2017,Schopf2018,Cavalazzi2021,Dodd2017}. A bacterium is a cell with the vital biological molecules
 in an aqueous environment protected by a cell membrane.  However, homochiral amino acids and polypeptides
in an aqueous environment, like in a living cell, are unstable, with racemisations of  L-amino acids within days to years,
depending on the physicochemical conditions \cite{Bada1972,Bada1985}, and 
enzymes within the lifespan of the living organism \cite{Fujii2010,Toxvaerd2025}.
 Since  vital molecules in a bacterium are unstable in the long term, a  possibility
is that they were originally formed elsewhere and in an environment where they were stable. 
  For could the complex and unstable structures of
the biomolecules be formed by gradual refinement in simple prebiological inorganic cells when these central
molecules  are unstable and, over time, lose their homochiral configurations? 

The Earth and the Solar system
 is $\approx$ 4.6 billion years old, and if life has originated in our solar system, these molecular structures and chemical reactions have been established in a
prebiotic aqueous environment, which differs from the aqueous environment in living organisms,
and in the time spent before the emergence of the first primitive living organisms.
 But that is actually all that can be said so far about when the asymmetric synthesis
 with the establishment of homochirality took place. 

 The answers to the two questions: where and how the homochiral structure was formed are connected. The homochirality in peptides is ensured by
 their secondary homochiral conformations with $\alpha-$helix and $\beta$-sheet conformations \cite{Pauling1951,Pauling1951a}. How the homochiral secondary
 structure was established can be answered with a physicochemical-based answer:
 by a spontaneous synthesis with negative reaction Gibbs free energy  $ \Delta G_r(aq)< 0$ \cite{Toxvaerd2017}.
This is, at first glance, just a formal thermodynamics clarification of the requirement for the life-giving synthesis, but it is also
the key to answering the two questions. For it must be a place with a lower water activity than in a biological cell. The aqueous environment in a biological cell can be
characterized as a diluted ionic aqueous solution with a suspension of biomolecules \cite{Bonucci2023,Tros2017},
i.e., with high water activity, which destabilizes the homochirality (see next chapter).
The Hadean ocean at the time before the first sign of life had presumably a concentration of salts  similar to the salt
concentrations in the oceans today \cite{Harrison2009,Harrison2020},
This salt concentration is significantly higher than the concentration in living cells \cite{Toxvaerd2023},
by which a spontaneous synthesis of homochiral peptides in an ocean-like environment is unlikely.
However, another fact that excludes the Hadean ocean as the place of emergence
of homochirality is that
the reaction Gibbs free energy for the inorganic synthesis of homochiral peptides in an ocean-like environment is positive due to the very low concentrations
of the reaction components, i.e., the concentrations of amino acids in the Hadean ocean. But an aqueous environment with low water activities 
is found in the Earth's crust  \cite{Yardley2009,Bucher2010,Stober2022}.

One theory for how homochiral molecules form is that they are catalyzed on stereospecific inorganic surfaces
\cite{Greber2006,Ernst2012,Rimola2013,Gellman2015,Buhse2021}. Such surfaces do indeed exist, and they are
undoubtedly also found at hydrothermal vents, where the earliest life forms have been observed \cite{Djokic2017,Cavalazzi2021,Dodd2017}.
But there is an overlooked problem: even if one has formed, for example, homochiral amino acids or homochiral peptides at a catalyzing surface,
they are unstable in a  biological aqueous environment. So the problem is not so much explaining how to
synthesize homochiral peptides. The problem is explaining how to maintain the
synthesized homochiral molecules over very long times, maybe for millions of years, in an aqueous environment to establish life.

In the article I argue that one can have stable polypeptides in an aqueous capillary environment over long periods,
which is necessary to ensure a desired stability of the homochirality of the building blocks of life near the place
where the origin of life may have occurred. In \cite{Toxvaerd2023,Toxvaerd2025} I formulated the hypothesis that our
life started in the Earth's crust with the synthesis of the vital molecules of life, and here I refine the hypothesis
by suggesting that it was formed in an aqueous capillary environment. The article is structured as follows: the next
section specifies the thermodynamic and general circumstances for an exergonic synthesis and long-term stability of
homochiral polypeptides, and in the following sections, what is known about the stability of proteins and what is known about the capillary environments in the crust. Finally, the hypothesis is specified together with possible ways to verify it.

\section{ Exergonic synthesis of homochiral peptides }
\subsection{The spontaneous polyperization of homochiral peptides: The Frank model}
In 1953, Frank published a theoretical model for spontaneous asymmetric synthesis and the emergence of homochirality.
Frank's model describes how a reaction network involving chiral species, autocatalysis, and inhibition can favor one enantiomer over another \cite{Plasson2004}.
The synthesis of homochiral peptides can be obtained by polymerization of chiral amino acids,
and the polymerization of peptides is an example of such a Frank synthesis with homochiral amplification. Numerous articles have
been published about asymmetrical polymerization of peptides with chiral amplification   \cite{Brandenburg2005,Ribo2017,Blanco2017,Buhse2022,Pineros2022,Shen2021}.
An important  and necessary condition for the network of consecutive reactions with homochirality of the product  is, as mentioned, the inclusion of
the racemization of the monomers. This is included in several of the reaction schemes in  the  articles, 
where the kinetic reactions
for the racemization of the chiral monomers are 
together with the polymerization of peptides, and the reaction equations with homochiral peptides
are solved for various values of reaction constants.

The  exergonic synthesis of homochiral peptides can be expressed by the sign of the reaction Gibbs free energy $\Delta G_r$
\begin{equation}
 \Delta G_r(aq)=\Delta H_r(aq) -T\Delta S_r(aq)< 0
\end{equation}	
where the reaction is the Frank-polymerization of a peptide $P$ from aqueous solutions of units of amino acids. The reaction can be simplified by the reaction scheme

\begin{figure}
\centering
\label{Figure 1.}
\includegraphics[width=0.80\linewidth]{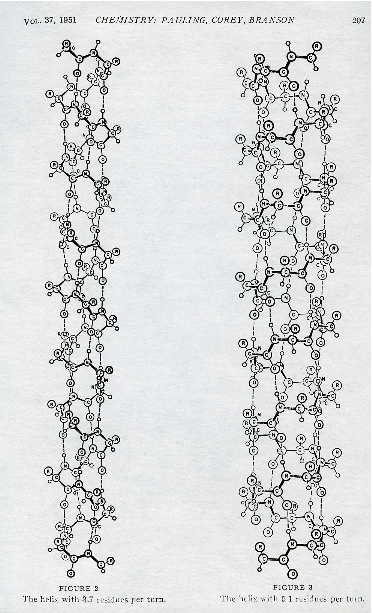}
\vspace{6pt}
\caption{Pauling's figures of the  secondary $\alpha$-helix  structure of a homochiral peptide chain \cite{Pauling1951}. The secondary structure is stabilized by
	weak hydrogen bonds. Each amide group is hydrogen-bound to the third (Pauling's Figure 2), or fifth (Pauling's Figure 3) amide group beyond
	it along the helix. The  $\alpha$-helix structure to the left of the figure (Pauling's Figure 2) is found in nature.}
\end{figure}

\begin{equation}
	\ce{\Sigma^n_{i=1,n} A_i(aq) ->[H_2O(aq),Cat] P_n(aq)},
\end{equation}
where the summation is over the amino acids $A_i(aq)$ in the peptide $P_n(aq)$ with $n$ units of amino acids, and the polymerization is catalyzed
by the aqueous environment ($aq$), and other factors (Cat).
Since the amino acids have an
 isomerization kinetics, the  Frank polymerization starts from a racemic mixture of the reactants $A_i(aq)$.

From a thermodynamics perspective, the details of the consecutive reactions are
unimportant, because the reaction Gibbs free energy 
 $ \Delta G_r(aq)$ is the difference between the state functions, and it depends solely  on the thermodynamic conditions of the reactants  $A_i(aq)$,
 and the final product, the homochiral peptide $P_n(aq)$.
 But in reality, the details, marked by $Cat$ in Eq. (1),  are important, and the reaction could, e.g., be catalysed at chiral surfaces or by
 amino acid salt bridges \cite{Meuzelaar2014}. The thermodynamic favorability of the overall reaction must be distinguished
 from the kinetic accessibility of peptide formation.

 The general thermodynamic condition for an exergonic synthesis to a product with a peptide with homochiral units is given in \cite{Toxvaerd2023}. It is due to
``chiral discrimination'' where the gain of enthalpy,  $ \Delta H_r(aq)< 0$,  exceeds the loss  of temperature times entropy
, $ -T\Delta S_r(aq)> 0$,  due to the ordered peptide structure, in the reaction Gibbs free energy  $ \Delta G_r(aq)=\Delta H_r(aq) -T\Delta S_r(aq)< 0$.
The reaction entropy $\Delta S_r(aq)$ is the difference between the entropy of $one$  peptide conformation of units of homochiral amino acids and the corresponding
entropy of the amino acids in an aqueous environment, and the reaction entropy is primarily determined by the low entropy of the ordered secondary conformation of the peptide.
So the sign of the reaction Gibbs free energy depends primarily on  the gain of enthalpy,  $ \Delta H_r(aq)< 0$, by creation of the
secondary conformation of the peptide with a sufficiently large negative enthalpy  $  H(P_n(aq))<< 0$ of the peptide.

(The successful spontaneous synthesis of homochiral peptides by the kinetics in the above-mentioned Frank models is achieved by choosing values
for the reaction constants in the consecutive reaction scheme that indirectly ensures a sufficiently negative
$ \Delta H_r(aq)$ for a homochiral peptide, and correspondingly, an unfavorable reaction enthalpy for a non-homochiral peptide. 
The kinetic reaction schemes operate with concentrations $[X]$
of the species $X$ in the consecutive reaction scheme, which are bulk concentrations in an infinite solvent, and they contain no details
about the stability of chemical bonding in the peptide and the sensitivity of the structure of the solvent.)

The  stability of homochiral peptides by a lare negative enthalpy  $ H(P_n(aq))$ 
is ensured by their secondary conformations and  is described by L. Pauling's  $\alpha-$helix and $\beta-$sheet structures 
\cite{Pauling1951,Pauling1951a}.

\subsection{Pauling's model for the secondary conformation of a peptide}

The homochirality of peptides is obtained and preserved by a synthesis of their
 secondary structure, first determined by Pauling \cite{Pauling1951}. Perhaps the figure of the $\alpha$-helix
 structure in Pauling et al's article best illustrates the condition for the  stability of the secundary conformation of peptide in an aqueous suspension.
The clockwise (positive) $\alpha$-helix structure (left) in Figure 1 with amide units of L-amino acids was deduced
from spectroscopic data for bond lengths and angles at the
planar peptide units, with coherent delocalized $\pi$ electrons along the chain of peptide bonds in the molecule,
 and with  $\it{intramolecular}$ stabilizing hydrogen-bonds  \cite{Arunan2011}.
The intramolecular hydrogen bonds are between a hydrogen atom at the substituted -N-H in a planar peptide bond with  an oxygen atom in a
nearby -C=O group in another peptide bond in the helix above or below. 
 These intramolecular bonds in the helix
can only be obtained with homochiral amino acids, by which the overall orientation changes with a constant 
angle $\approx - 57^{\circ}$  at each covalent bond in the helix. This angle at the peptide bond  is changed  to  $\approx 57^{\circ}$ if an L-configuration
is changed to a D-configuration,  the hydrogen bond is broken, and the 
helix structure is destabilized. This fact is confirmed by the observation that a change of an L-structure to a D-structure in one chiral unit
is associated with a loss of the secondary structure of the peptide and an increased concentration of water molecules in a peptide \cite{Fujii2010}.

\subsection{The tertiary structure of proteins}

The energy of a -N-H$\cdot\cdot\cdot$O=C- hydrogen bond is only a fraction of
 the energy of a
hydrogen bond between oxygen and hydrogen in pure water \cite{Deshmukh2009,Sun2010,Wendler2010,Dabrowski2020}, and the stabilizing intramolecular hydrogen bonds compete with
hydrogen bonds between the  -N-H groups and oxygens in the water molecules in the solvent.  The 
 -N-H groups in the helix will have a higher tendency to form intermolecular hydrogen bonds
 with the water molecules if the protein is surrounded by pure water,   and the $\alpha-$helix structure might be
 destabilized in peptide suspensions in pure water, and the protein in a pure water suspension can, over time, lose its helix structure. However, 
the secondary structure will be more stable if the water activity is reduced, e.g., by ions Na$^+$, K$^+$, Ca$^{2+}$,..., or near hydrophilic surfaces such as silica surfaces
in capillary environments in the Earth's crust \cite{Ilgen2023}.

The conformational  folding and the  stability of a protein
is determined by  its  tertiary  conformation  in an aqueous environment \cite{Gadzala2019}, where
the thermodynamic force, surface tension, also co-determines the conformation.
Proteins are normally compact globular molecules, where $\alpha$-helices and $\beta$-sheets are packed into a compact molecule.
The compact proteins contain a small amount of units of D-amino acids \cite{Armstrong2023}, and  the achiral glycine and  D-amino acids
make the protein more flexible and able to adopt a compact globular conformation.

While amino acids are readily soluble, peptides are not, and proteins in a cell can be described as an aqueous emulsion
of compact proteins, which is stabilized by the surface
tension between the aqueous suspension and the lipophilic protein, and intramolecular forces. However,
some proteins have long hydrophilic chains, where the hydrophilicity is achieved
by units with charges \cite{Kragelund2024}. Milk is an example of such proteins that form an emulsion of micelles.

Overall, the compact conformation of a protein with
$\alpha$-helices and $\beta$-sheets is ensured by chains of delocalized $\pi$-orbitals, by intramolecular hydrogen bonds, and by a surface tension to the surrounding water.

\subsection{ Prebiotic exergonic synthesis of homochiral peptides }

The overwhelming information on the \textit{ in vivo} structure of proteins shows us that they are much more stable in terms
of homochirality than their building blocks, the amino acids. Peptides and proteins are fairly stable over the lifetime
of a living organism. So the kinetic models for Frank synthesis of peptides correctly describe how the asymmetric synthesis
can spontaneously ensure a homochiral peptide from a racemic mixture of amino acids by a synthesis to Pauling secondary conformation 
 in a stable energy minimum of the  Gibbs free energy.

From a Darwinian perspective, a protein must be stable long enough to ensure that an individual living system can form a new individual with
 perfect proteins, and it is not required that the protein be in a conformation with a $global$ Gibbs free energy minimum \cite{Anfinsen1973}.
It is good enough to be in a local minimum, as is often the case, and
 proteins with $\alpha$-helix and $\beta$-sheet conformations  are often found  in local  energy minima \cite{Yang2012}.
Nature also abounds with examples of proteins with open peptide chains \cite{Kragelund2024},
and small peptides with several secondary conformations with local free energy minima
\cite{Mitsuta2019}.
But the situation is different if the start of life took millions of years with a slow synthesis of its building blocks.
Here, long-term stability of the homochiral conformations in an aqueous environment is necessary.

\section{Stability of homochiral peptides}

There are countless suggestions as to how homochirality in peptides an  in living organisms arose, e.g.:
whether homochirality came to Earth with homchiral molecules in meteors, whether it was catalyzed by wet and chiral surfaces,
or whether it simply was there at the origin of life. But it is, in fact, irrelevant  since 
homochiral proteins  in biosystems are unstable. Some examples:

An example of proteins being unstable and losing their homochirality is cataracts,
which develop in older people. The disease is due to changes in the proteins in the lens of the eye.
The lenses with their contents of homochiral proteins are formed during  the  pregnancy, and then they
are without metabolism. Old lenses contain D-amino acids, and lenses  with cataract contain
an excess amount of D-Aspartic acid  \cite{Hooi2011,Hooi2013},
and it is due to a destruction of the homochiral proteins and a racemization of the amino acids.

Another example of protein instability is seen in Parkinson's and Alzheimer's diseases.
The diseases most often occur in older people, and now that other causes of death, such as cardiovascular disease
and cancer, can be cured to a certain extent, diseases that can be attributed to defects in the long-lived nerve cells are becoming more dominant.
As with cataracts, D-amino acids can be detected  in patients suffering from Parkinson's disease \cite{Gavasoni2025} and
Alzheimer's disease \cite{Liu2023}. The appearance of D-amino acids in humans can be used as biomarkers for neurological disorders \cite{Talebi2026}.

The instability of homochiral proteins  with the loss of their
secondary conformations are not limited to humans, but is found in all living organisms from the simplest form of life, the  bacteria \cite{Nystrom2003,
Nystrom2007,Wagener2025}.

Some theories of the origin of life assume that it was genetics that ensured homochirality. However, the backbones of DNA
and RNA are homochiral sugar molecules whose homochirality is ensured enzymatically \cite{Toxvaerd2018}. So proteins with their homochirality
are the central homochiral molecules in life's network of biological reactions.

A living organism ages and dies due to many factors, but if its central building blocks, the proteins, are unstable, it dies sooner or later.
If a living system can not maintain the homochirality in proteins, the central questions are not only how homochirality was ensured at the origin of life but also:\\
\textit{How the homochiral proteins, which appear in the simplest living organisms, could maintain their homochiral form
in a prebiotic environment over very long periods of time preceding the formation of these simplest cells?}

However, the instability of proteins in living organisms also provides a hypothesis for how  and where the long-term stability
of peptides was ensured at the beginning of life. This could have happened in an aqueous prebiotic environment, where
homochiral peptides were more stable than they are in the aqueous environment in living cells. The water activity in
living cells is quite high and resembles a dilute electrolyte solution \cite{Tros2017,Lang2023}, and the high water activity undermines the stability of the
secondary conformation in the peptides. But a more stable aqueous
 protein-stable environment could be a capillary environment with lower water activity
than in living cells, in cracks and fissures in the vents of the Earth's mantle.

\section{Emergence of homochiral peptides in Earth's crust}
If one imagines that the prebiotic synthesis of the building blocks of life might have taken millions of years,
one must demand that homochiral building blocks be very stable. This means that we must search for environments
and circumstances where not only can homochiral peptides be synthesized, but they are also long-term stable.
 Such a place could be the capillary environment in the Earth's crust.

\subsection{The Earth's crust}
Earth's crust is the outer shell of rocks in the lithosphere, the solidified division of Earth's top layers that includes the crust and the upper part of the mantle.
The crust is divided into the tectonic plates with
    the continental crust 25–70 km, which is  thick and mostly composed of less dense, more felsic rocks, such as granite, and the
    oceanic crust 5–10 km thick, which is composed primarily of basalt.
The temperature of the crust increases with depth, reaching values typically in the range from about 700 to 1,600 $^{\circ}$C
 at the boundary with the underlying mantle. The temperature of the crust and underlying mantle is estimated to be similar or only slightly higher in the past at the 
 origin of life \cite{Korenaga2026}.

Earth formed approximately 4.6 billion years ago, and the oldest continental crustal rocks on Earth today were created in the \textit{Primordial Earth Eon}
in the range from about 3.7 to 4.28 billion years,
and are located in Western Australia, in the Canadian Shield, and on other cratonic regions, such as those on the Fennoscandian
Shield.  Earth has been forming a secondary and tertiary crust, and the average age of
Earth's current continental crust has been estimated to be about 2.0 billion years.
The geophysics evolution of the Earth's continents is reviewed in \cite{Cawood2022}.
 
\subsection{Capillary environment  in the crust}
The Earth's mantle and crust contain large amounts of water, probably exceeding the total
amount of water in the Earth's oceans. Estimates of the amount of water range from 1-12 ``ocean masses'' \cite{Ohtani2021},
and the latest reported study suggests that there is at least the equivalent of 3 oceans of water in the Earth's mantle and crust\cite{Ren2025}.
The water storage in the crust and mantle is connected, and with the exchange of water \cite{Ohtani2024}, and recent results suggest that
the water in the crust and mantle plays a significant role in the drift of the tectonic plates \cite{Quin2025}.

 The onset of the Earth's hydrological cycle appered four billion years ago or earier \cite{Gamaleldien2024}.
 The occurrence of water and salinity in the Earth's crust is reviewed in \cite{Yardley2014}. Water has played a crucial role in the 
evolution of the continental crust \cite{Yardley2009,Bons2012}. From deep drilling in the continental crust
\cite{Stober2022} and geological studies, it can be concluded that water occurs in fractures of an interconnected network of
water-conducting vents in the crust \cite{Bucher2010}.  
This is also where the earliest biological traces 3.5-3.8 billion years old are found  \cite{Djokic2017,Schopf2018,Cavalazzi2021,Dodd2017}.

\subsection{Synthesis and conservation of life's prebiotic building blocks in the vents  }
One of the many theories about the formation of homochirality in biological systems is that it occurred
by an asymmetric synthesis at chiral crystalline surfaces \cite{Zhao2020,Xu2021}. Although the central problem is not the
formation but the preservation of prebiotic homochiral molecules
over very long times, it is also important whether
the synthesis actually occurred by catalysis of chiral crystalline surfaces. An asymmetric spontaneous synthesis of racemic amino acids to peptides with the
secondary homochiral conformation is to a low enthalpic state, and it is very unlikely without stereospecific catalysis.  
Such chiral crystalline surfaces are indeed found in the Earth's crust \cite{Hazen2010,Ma2017,Avnir2024},
and if the emergence of peptide homochirality was in hydrophilic vents in the crust, it must be
 likely that homochiral peptides with the secondary conformation were synthesized at chiral hydrophilic
silica surfaces in water-filled vents in the Earth's crust.

Silicate surfaces are generally very hydrophilic, and microcrystalline silicate materials such as
clay are highly water-absorbent and remain wetted even at low water vapor pressures.
Some mineral surfaces, including silica surfaces, can even be described as
superhydrophilic with prewetting of the crystal surface for low vapor pressures \cite{Drelich2011}, and with enantioselective adsorption \cite{Huang2024a}.
According to the previous chapter, this means that 
 an adsorbed peptide with a secondary homochiral conformation are stabilized, and that the  water molecules do not easily destabilize the peptide.

 \section{The hypothesis: Peptides are stable in capillary aqueous environments}
This article proposes that is that not only were the biological homochiral peptides synthesized
in a capillary aqueous environment in vents in the Earth's crust, but they were also long-term stable
in the capillary environment, unlike peptides in living cells, and this long-term stability enabled enzymatic inorganic
synthesis of other chiral biomolecules at the origin of life.

\subsection{A posible verification of the origin of homochirality}
There are many examples of asymmetric synthesis of homochiral molecules, but no examples of a direct synthesis of peptides
with the homochiral secondary conformation. If possible, one should probably try to synthesize peptides in a capillary
water-saturated environment, but with low water activity (i.e., high ionic concentration), and at chiral crystalline surfaces.
Such a successful experiment will be a crucial support to the hypothesis.

A less ambitious experimental study would be to determine the stability of a protein's secondary conformation as a function of the water activity.
The water activity in living cells is similar to  a ``physiological salt solution'', and one would expect the protein's secondary conformation
to be more stable in a lower water activity and correspondingly more unstable in e.g. pure water.
One shall also expect the protein to be more stable in a capillary environment, e.g., in an aqueous slurry of clay.
With the experience from the lifespan of living organisms, even such an experiment about the stability of homochirality
can take a very long time.  But proteins are destabilized by an increased temperature, and one shall expect that the
destabilization is more pronounced at high water activities. By extending the study by
investigating the stability of a protein as a function of temperature and the aqueous activity,
it might be possible to prove that proteins in a capillary environment 
are stable contrary to a pure aqueous environment.

$   $ \\


\begin{thebibliography}{99}
	
\bibitem[Djokic(2017)]{Djokic2017}Djokic, T., Van Kranendonk, M.J., Campbell, K.A., Walter, M.R.,
	Ward, C.R., 2017. Earliest signs of life on land preserved in ca. 3.5 Ga hot spring deposits.  Nat. Commun. 8:15263.
\bibitem[Schopf(2018)]{Schopf2018}Schopf, J.W., Kitajima, K., Spicuzza M.J., Kudryavtsev, A.B., Valley 
	J.W.,  2018.  SIMS analyses of the oldest known assemblage of microfossils document their taxon-correlated carbon isotope compositions.
		Proc. Nat. Acad. Sci. (U.S.A.) 115:53-58. 
\bibitem[Cavalazzi(2021)] {Cavalazzi2021} Cavalazzi, B., Lemelle, L., Simionovici, A.,   et al., 2021.
	Cellular remains in a $\approx$ 3.42-billion-year-old subseafloor hydrothermal environment.
		Sci. Adv. 7:eabf3963.
\bibitem[Dodd(2017)]{Dodd2017} Dodd, M.S., Papineau, D., Grenne, T., Slack, J.F.,  et al, 2017.
	Evidence for early life in Earth's oldest hydrothermal vent precipitates.
	Nature  543, 60-74.
\bibitem[Bada(1972)]{Bada1972} Bada, J.L., 1972.
	Kinetics of Razimization of Amino Acids as a Function of pH.  J. Am. Chem. Soc. 94, 1371--1373.	
\bibitem[Bada(1985)]{Bada1985} Bada, J.L., 1985. Racemization of Amino Acids in G. C. Barrett(ed.) Chemistry and Biochemistry of Amino Acids (Chapman and Hall 1985).
\bibitem[Fujii(2010)]{Fujii2010} Fujii. N., Fijii, N., Kida, M., Kinouchi, T., 2010.
	Influence of L$\beta$-, D$\alpha$-Asp isomers of the Asp-76 residue of
	the properties of $\alpha$A-crystallin 70-88 peptide.  Amino Acids 39, 1393-1399.  
\bibitem{Toxvaerd2025} Toxvaerd, S., 2025. Origin of homochirality in peptides: The first milestone at the origin of life.
	BioSystems 253, 105479. 
\bibitem[Pauling(1951)]{Pauling1951} Pauling, L., Corey, R.B., Branson, H.R., 1951. The structure of proteins:
	Two hydrogen-bonded helical configurations of the polypeptide chain. Proc. Natl. Acad. Sci. (U.S.A.) 37, 205-211. 
\bibitem[Pauling(1951a)]{Pauling1951a} Pauling, L., Corey, R.B., 1951. Configurations of polypeptide chains with favored orientations around single bonds:
	two new pleated sheets. Proc. Natl. Acad. Sci. (U.S.A.)   37, 729-740. 
\bibitem[Toxvaerd(2017)]{Toxvaerd2017} Toxvaerd, S., 2017.
	The role of the peptides at the origin of life.  J. Theor. Biol. 429, 164-169. 
\bibitem{Bonucci2023} Bonucci, M., Shu, T., Holt, L.J., 2023. How it feels in a cell.  Trends Cell Biol. 33, 924-938. 
\bibitem[Tros(2017)]{Tros2017} Tros. M., Zheng, L., Hunger, J., Bonn, M., Bonn, D.,  et al, 2017.
	Picosecond orientational dynamics of water in living cells.
	Nature Commun. 8, 904.
\bibitem[Harrison(2009)]{Harrison2009} Harrison, T.M., 2009.  The Hadean Crust:
	Evidence from  $>$4 Ga Zircons. Annu. Rev.  Earth  Planet. Sci.   37, 479-505.
\bibitem{Harrison2020} Harrison T.M., 2020.   Hadean Earth. Springer, Cham, Switzerland,ISBN: 978-3-030-46687-9, 2020.
     \bibitem{Toxvaerd2023} Toxvaerd, S., 2023. Origin of Homochirality: The Formation and Stability of Homochirale Peptides in Aqueous
	     Prebiological Environment in the Earth's Crust. Symmetry 15, 155.
\bibitem{Yardley2009}  Yardley, B.W.D., 2009. The role of water in the evolution of the continental crust. J. Geol. Soc. London 166, 585-600.
\bibitem{Bucher2010} Bucher, K., Stober, I., 2010. Fluids in the upper continental crust.
	Geofluids 10, 241-253.
\bibitem{Stober2022} Stober, I., Giovanoli, F., Wiebe, V., Bucher, K., 2022.  Deep hydrochemical section through the Central Alps:
	evolution of deep water in the continental upper crust and solute acquisition during water-rock interaction.
		 J. Geosci 115, 9.	
\bibitem{Greber2006} Gerber, T., \u{S}ljivan\u{c}anin, \u{Z}., Schillinger, R., Wider, J., Hammer, B., 2006.
	Chiral Recognition if Organic Molecules by Atomic Kinks on Surfaces. Phys. Rev. Lett. 96, 056103. 
\bibitem{Ernst2012} Ernst, K.-H., 2012. Molecular chirality at surfaces. Phys. Status Solidi B 2012, 249, 2057-2099.
\bibitem{Rimola2013}Rimola, A., Costa, D., Sodupe, M., Lambert, J.-F., Ugliengo, P., 2013. Silica Surface Features and Their Role
	in the Adsorption of Biomolecules: Computational Modeling and Experiments. Chem. Rev. 113, 4216-4313. 
\bibitem{Gellman2015} Gellman, A.J., Tysoe, W.T., Zaera, F., 2015. Surface Chemistry for Enantiodelective   Catalysis.  Catal. Lett. 145, 220-232.
\bibitem{Buhse2021} Buhse, T., Cruz, J.-M., Noble-Te\'{a}n, M., Hochberg, D., Rib\'{o}, M., Crusats, J., Micheau, J.-C., 2021.
	Spontaneous Deracemizations.  Chem. Rev. 121, 2147-2229. 
\bibitem[Frank(1953)]{Frank1953} Frank, F.C., 1953. On spontaneous asymmetric Synthesis. Biochin. Biophys. Acta  11, 459-463. 
\bibitem[Plasson(2004)]{Plasson2004} Plasson, R., Bersinni, H., Commeyras, A., 2004. Recycling Frank:
	Spontaneous emergence of homochirality in noncatalytic systems. Proc. Nat. Acad. Sci. (U.S.A.)    101  , 16733-16738. 
\bibitem{Brandenburg2005} Brandenburg, A., Andersen, A. C., H\"{o}fner, S., Nilsson, M., 2005.
	Homochiral growth through enantiomeric cross-inhibition.
	 Origin Life Evol. Biosphere  35, 225-241. 
\bibitem{Ribo2017} Rib\'{o}, J.M., Crusats, J., El-Hachemi, Z., Moyano, A., Hochberg, D., 2017.
	Spontaneous mirror symmetry breaking in heterocatalytically coupled
	enantioselectivve replicators.  Chem. Sci. 8, 763-769.
\bibitem[Blanco(2017)]{Blanco2017} Blanco, C.,  Stich, M.,  Hochberg, D., 2017.  Mechanically Induced Homochirality in Nucleated Enantioselective
	Polymerization.  J. Phys. Chem. B 121, 942-955.  
\bibitem{Buhse2022} Buhse, T., Micheau, J.-C., 2022. Spontaneous Emergence of Transient Chirality in Closed, Reversible Frank-like Deterministic Models.
	Origin Life Evol. Biosphere 52, 3-20.
\bibitem{Pineros2022} Pi\~{n}eros, W.D., Tlusty, T., 2022.  Spontaneous Chiral symmetry breaking in a random driven chemical system.
	 Nat. Commun. 13, 224. 
\bibitem[Shen(2021)]{Shen2021} Shen, Y., Wang, Y., Hamley, I.W., Qi, W., Su, R., He, Z., 2021.
Chiral self-assembly of peptides: Toward the design of supramolecular polymers with enhanced chemical and biological functions.
Prog. Polym. Sci.  123, 101469.
\bibitem{Meuzelaar2014} Mauzelaar, H., Tros, M., Huerta-Viga, A., van Dijk, C.N., Vreede, J., Woutersen, S., 2014.
	Solvent-Exposed Salt Bridges Influence the
	Kinetics of $\alpha-$Helix Folding and Unfolding. J. Phys. Chem. Lett.  5, 900-904. 
 \bibitem{Arunan2011}Arunan, E., Desiraju, G.R., Klein, R.A., Sadlej, J., et al, 2011. Defining the hydrogen bond: An account (IUPAC Technical Raport)$^*$.
	Pure Appl. Chem. 83, 1619-1636.
\bibitem{Deshmukh2009} Deshmukh, M.M., Gadre, S.R., 2009. Estimation of N-H....O=C Intramolecular Hydrogen Bond Energy in Polypeptides. 
	J. Phys. Chem. A 113, 7927-7932.
\bibitem{Sun2010} SUN. C.-L., Wang, C.-S., 2010. Estimation on the intramolecular hydrogen-bonding energies
	in proteins and peptides bythe analytic potential energy function. THEOCHEM 956, 38-43.
\bibitem[Wendler(2010)]{Wendler2010} Wendler, K., Thar, J., Zahn, S., Kirchner, B., 2010. Estimating the Hydrogen Bond Energy.
	J. Phys. Chem. A 114, 9529-9536.  
\bibitem{Dabrowski2020}Dabrowski, J., Nowak, W., Ptak, A., 2020. How strong are hydroden bond in the peptide model? PCCP 22, 1392.
\bibitem{Ilgen2023} Ilgen, A.G., Senanayake, H.S., Thompson, W.H., Geathouse, J.A., 2023
Structure and energetics of hydrogen bonding networks in dilute	HOD/H$_2$O solutons confined in silicananopores. Environ. Sci. Nano 10, 3025-3038.
\bibitem{Gadzala2019} Gadza\l a, M., Du\l ak, D., Kalinowska, B., Bastar, Z., et al., 2019.
	The aqueous environment as an active participant in the protein folding process. J. Mol. Graph. Model. 87, 227-239.  
\bibitem{Armstrong2023} Armstrong, D., Berthod, A., 2023. Occurrence of D-amino acids in natural products. Nat. Prod. Bioprospect 13, 47. 
\bibitem{Kragelund2024} Holehouse, A.S.,  Kragelund, B.B., 2024. The molecular basis for cellular function of intrinsically disordered protein regions.
Nat. Rev. Cell Biol. 25, 187-211.
\bibitem{Anfinsen1973} Anfinsen, C.B., 1973. Principles that Govern the Folding of a Protein Chains. Science 181, 223-230.
\bibitem{Yang2012} Yang, L.-Q., Ji, X.-L., Liu, S.-Q., 2012. The free energy landscape of protein folding and dynamics: a global view.
J. Biol. Struc. Dyn. 31, 982-992.
\bibitem{Mitsuta2019} Mitsuta, Y., K\"{a}stner, J., Yamanaka, S., Kawakami, T., Okumura, M., 2019. Free energy reaction root mapping of alanine tripeptide in water.
Mol. Phys. 117, 2284-2292.
\bibitem{Hooi2011} Hooi, M.Y.S., Truscott, R.J.W., 2011. Racemisation and human cataract. D-Ser,
	D-Asp/Asn and D-Thr are higher in the lifelong proteins of cataract lenses than in age-matched normal lenses. AGE 33, 131-141.
\bibitem{Hooi2013} Hooi, M.Y.S., Raftery, M.J., Truscott, R.J.W., 2013.
	Age-dependent racemization of serine residues in a human chaperone protein. Protein Sci. 32, 93-100.
\bibitem{Gavasoni2025} Gervasoni, J., Marino, C., Imarisio, A., Santussi, L., et al, 2025.  \textit{et al.}. Independent serum metabolomics approaches
	identify disrupted glutamic acid and serine metabolism in Parkinson's disease patients. npj Parkinson's disease 11, 274 
\bibitem{Liu2023} Liu, M., Li, M., He, J., He, Y., Yang, J., Sun, Z., 2023.
	Chiral Amino Acid Profiling in Serum Reveals Potential Biomarkers for Alzheimer's Disease.  
	J. Alzheimer's Dis. 94, 291-301. 
\bibitem{Talebi2026} Talebi, M., Sharifi-Ardani, S. E., Ghannadzadeh, F. S., Sattarpour, M., et al., 2026.
	Detection and functional assessment of D-amino acids as a potential biomarkers in neurological disorders. Mol. Bio. Reports 53, 60.
\bibitem{Nystrom2003}Nystr\"{o}m, T., 2003. Conditional senescence in bacteria:death of the immortals. Mol. Microbiol. 48, 17-23. 	
\bibitem{Nystrom2007}Nystr\"{o}m, T., 2007. A Bacterial Kind of Aging. Plos Genetics 3, e224.
\bibitem{Wagener2025}Wagener, L., Nath, A., Tu\u{g}rul, M., Aertsen, A., Steiner, U.K., Proenca, A.M., 2025. Protein aggregation drives cell aging in a size-specific 
	manner in \textit{ Escherichia coli}. Mbio, mbio.02050-25.
\bibitem{Toxvaerd2018}	Toxvaerd, S., 2018, The start of the Abiogenesis: Preservation of homochirality in proteins as a necessary and
sufficient condition for the establishment of the metabolism.
 J. Theor. Biol. 451, 117-121.
\bibitem{Lang2023} Lang, X., Shi, L., Zhao, Z., Min, W., 2023. Probing the structure of water in individual living cells.
	Nat. Commun. 15, 5271.
\bibitem{Korenaga2026} Korenaga, J., Hertzberg, C., 2026. Thermal history of Earth's mantle: Implications for Earth system evolution and lessons from recent studies.
Earth Planet. Sci. Lett. 683, 119958.
\bibitem{Cawood2022} Cawood, P.A., Chowdhury, P., Mulder, J.A., Hawkesworth, C.J., et al., 2022. Secular Evolution of Continents and the Earth System.
Review of Geophysics 60, e2022RG000789.
\bibitem{Ohtani2021} Ohtani, E., 2021. Hydration and Dehydration in Earth's Interior. Annu. Rev. Earth Planet. sci. 49, 253-278.
\bibitem{Ren2025} Ren, Z.Y., Xie, Y.F., Chen, C.J., Yao, H.B., Tang, J.T., Zhang, K., 2025. New inasights into Earth's mantle conductivity and water distribution
	using Macau Science Satelite-1 data. Earth and Planetary Phys. 9, 595-606.
\bibitem{Ohtani2024} Ohtani, E., Ishii, T., 2024. Role of water in dynamics of slabs and surrounding mantle. Prog. earth and planet. sci. 11, 65. 
\bibitem{Quin2025} Quin, S.-P., Gazel, E., Wang, J.-H., 2025. Mantle transition zone water triggers lithospheric weakening and spreading.
	Geology 53, 461-466. 
\bibitem{Gamaleldien2024}  Gamaleldien, H., Wu, L.-G., Olierook, H.K.H., Kirkland, C.L., et al., 2024. Onset of the Earth's hydrological cycle
	four billion years ago or earlier.  Nature geoscience  17, 560-565.
\bibitem{Yardley2014} Yardley, B.W.D., Bodnar, R.I., 2014. Fluids in the Continental Crust. Geochem. Perspectives 3, 1-127. 
\bibitem{Bons2012} Bons, P.D., Elburg, M.A., Gomez-Rivas, E., 2012. A review of the formation of tectonic veins and their microstructures.
	J. Struc. Geol. 43,33-62.
\bibitem{Zhao2020} Zhao. X., Zang, S.-Q., Chen, X., 2020. Stereospecific interactions between chiral inorganic nanomaterials and biological systems.
	Chem. Soc. Rev. 49, 2481-2503.
\bibitem{Xu2021} Xu, Y., Duan, J.-J., Yi, Z.-Y., Zhang, K.-X., Chen, T., Wang, D., 2021. Chirality of molecular nanostructures
	on surfaces via molecular assembly and reaction: manifestation and control. Sur. Sci. Peports 76, 100531.
\bibitem{Hazen2010} Hazen, R.M., Sverjensky, D.A., 2010. Mineral Surfaces, Geochemical Complexities, and the Origin of Life.
 Cold Spring Harb. Perspect. Biol. 2, a002162.
\bibitem{Ma2017} Ma, W., Xu, L., de Moura, A.F., Wu, X., et al., 2017. Chiral Inorganic Nanostructures. 
 CHem. Rev. 117, 8041-8093.
\bibitem{Avnir2024} Avnir, D., 2024, Chiral Minerals. Minerals 14, 995.
\bibitem{Drelich2011} Drelich, J., Chibowski, E., Meng, D.D., 2011. Terpilowski K. Hydrophilic and superhydrophilic suraces
	and materials.  Soft Matter 7, 9804-9828.
\bibitem{Huang2024a} Huang, T.-Y., Sun, J., Wang, D.-X., Wang, Q.-Q., 2024. Recent progress in chiral zeolites: Structure,
	synthesis, characterization and application. Chinese Chem. Lett. 35, 109758.
\end{thebibliography}
\end{document}